\documentclass[
    ,final            
  ]
  {aipproc}

\layoutstyle{6x9}

\begin{document}

\title{Anti-ferrodistortive Nanodomains in PMN Relaxor}

\author{A. Tkachuk}{
  address={Department of Materials Science and Engineering and Materials Research Laboratory,
 University of Illinois at Urbana-Champaign, Urbana, IL 61801}
 ,altaddress= {Current addresses: Advanced Photon Source, Argonne National Laboratory, Argonne, IL 60439}
}
 
\author{Haydn Chen}{
  address={Department of Materials Science and Engineering and Materials Research Laboratory,
 University of Illinois at Urbana-Champaign, Urbana, IL 61801}
,altaddress={Department of Physics and Materials Science, City University of Hong Kong, Kowloon}
}


\begin{abstract}
 Temperature dependent studies of the 1/2(hk0) superlattice reflections ($\alpha$ spots) by synchrotron x-ray scattering measurements were performed in Pb(Mg$_{1/3}$Nb$_{2/3}$)O${_3}$ (PMN) and [PbMg$_{1/3}$Nb$_{2/3}$O$_{3}$]$_{1-x}-$ [PbTiO3]$_{x}$ \ (PMN-xPT) with Ti doping $x\leq0.32$ single crystals. Separation of the $\alpha$ spots from the underlying diffuse scattering background allowed studying them as separate entities for the first time. Structure factor calculations have shown that $\alpha$ spots constitute the presence of a new kind of anti-ferrodistortive nanoregions (AFR) in the form of fluctuations produced by anti-parallel short-range correlated $\langle110\rangle$ Pb$^{2+}$ displacements. AFR appear to be different and unrelated to the chemical nanodomains (CND) and ferroelectric polar nanoregions (PNR). Simultaneous presence of AFR and PNR can explain relaxor behavior as a result of competition between randomly occurring ferroelectric and anti-ferroelectric fluctuations. Temperature dependence of the \ensuremath{\alpha} spots in PMN showed a direct correlation with the freezing phase transition near T$_{f}{\approx}$220 K.
 
\end{abstract}

\maketitle


\section{Introduction}

	Pb(Mg$_{1/3}$Nb$_{2/3}$)O${_3}$  (PMN)  \cite{AVT.45} is a relaxor ferroelectric, which among other properties exhibits strong dependence of real $\chi'$ and imaginary $\chi''$ parts of dielectric susceptibility on temperature and frequency $f$ of the applied AC electric field. A broad peak in $\chi'$ temperature dependence, for example,  occurs at 265~K ($f$=1~kHz), which shifts to $T_{f}\approx$220~K  as  $f$ decreases to zero  \cite{AVT.45}.  

Although average crystal structure is indistinguishable from cubic at all temperatures (5-800 K)  \cite{AVT.51, AVT.52}, ferroelectric rhombohedral polar nanoregions (PNR) were postulated to exist in PMN below T$_{d}\approx$620 K  \cite{AVT.218}. Relaxor behavior in PMN is commonly attributed to PNRs, which undergo cooperative freezing near T$_{f}$ 
into a glass-like state, reminiscent of magnetic spin glasses, due to random fields produced by underlying chemical and displacement disorder \cite{AVT.45, AVT.49, AVT.48, AVT.168}. Spherical-Random-Bond-Random-Field (SRBRF) model is a recent self-consistent theoretical treatment for PMN type relaxors as a special kind of spherical dipole glass \cite{AVT.168}.

The short-range scale of the nanodomains complicates the direct observation of the PNRs with diffraction techniques. Nevertheless, existence of the PNRs as ferroelectric fluctuations was attributed to the strong temperature dependent neutron diffuse scattering near the Brillouin zone center  ($q\ll$0.1 r.l.u.) below T$_{d}$ \cite{AVT.58, AVT.67}. Correlation radius of these fluctuations was shown to increase from \ensuremath{\sim}50 {\AA} near T$_{d}$ to \ensuremath{\sim}200 {\AA} saturation value below T$_{f}$ \cite{AVT.58}. Alternatively, the anisotropic x-ray diffuse scattering along $\langle01\bar{1}\rangle^*$ cubic reciprocal lattice directions was attributed to pure transverse optical (TO) soft modes \cite{AVT.25}. Other workers claimed to resolve diffuse scattering interpretation controversy by introducing a phase-shifted soft mode model of PNRs \cite{AVT.1000}. Moreover, inelastic neutron scattering experiments \cite{AVT.11, AVT.213} identified the TO soft mode above T$_{d}$, which becomes overdamped at lower temperatures due to condensation of the aforementioned PNRs \cite{AVT.213}. Interestingly, subsequent recovery of the Curie-Weiss behavior coincides with the freezing temperature T$_{f}$ \cite{AVT.1001}.  The presence of a distinct thermodynamic phase transition near T$_{f}$ into a nonergodic frozen dipolar glass state was suggested in the past based on electroacoustic studies of PMN \cite{AVT.137}.

An evidence of a new kind of anti-ferrodistortive nanoregions (AFR) in PMN, which can be envisioned as fluctuations different from PNRs, is presented in this work based on temperature dependence and structure factor calculations of 1/2(hk0) short-range order superlattice peaks. The simultaneous presence of AFRs and PNRs is important for understanding PMN relaxor in terms of competing anti-ferroelectric and ferroelectric interactions on the nanometer scale, which is supported by the SRBRF model \cite{AVT.168}.
   
\section{Experimental}

All scattering studies were performed on PMN single crystals produced by Czochralski and Bridgeman methods. PMN crystals doped with PbTiO$_{3}$ (PMN-xPT) x\ensuremath{\leq}0.32 were grown by the melted flux method. Crystals were in the form of  $\langle001\rangle$ or $\langle111\rangle$ oriented platelets having surfaces with linear dimensions no larger than 3-7 mm and thickness ${\sim}$1 mm. PMN(111) crystals were sputtered with gold for in-situ x-ray measurements under applied electric field (up to 4 kV/cm). All crystals were of good quality with a mosaic spread no worse than 0.01\ensuremath{^\circ}, except PMN-0.32PT, which exhibited \ensuremath{\sim}0.3\ensuremath{^\circ} mosaic, obtained from x-ray diffraction rocking curve measurements. Dielectric spectroscopy results obtained from the same crystals were published previously \cite{AVT.114}.

	Synchrotron x-ray work was conducted on X-18A beamline at the National Synchrotron Light Source (NSLS), Brookhaven National Laboratory, and at 33-ID beamline at the Advanced Photon Source (APS), Argonne National Laboratory. Both beamlines used focusing mirrors, which also served as high energy harmonic discriminators. Crystals were studied in 10-300 K range inside closed cycle He gas cryostats mounted on 4-circle or 6-circle kappa diffractometers.
	
	Measurements in 300-800 K range were conducted on in-house (CuK$_{\ensuremath{\alpha}}$, 40 kV, 200~mA) Rigaku rotating anode source equipped with focusing pyrolytic graphite monochromator located before the sample mounted inside the evacuated  heating stage.

	Energy of the incident x-rays at NSLS was chosen to be 10 keV for optimal beamline performance. At APS the x-ray energy in addition to 10 keV was tuned near Nb ${K}$ (18.99 keV) and Pb $L_{III}$ (13.035 keV). The size of the incident beam was collimated by a pair of slits set to 0.5 $\times$ 0.5 mm$^{2}$ at APS, 1 $\times$ 1 mm$^{2}$ at NSLS and 2 $\times$ 2~mm$^{2}$ for rotating anode. Diffuse scattering measurements at rotating anode required open detector slits and 40 sec/point counting times to achieve reasonable statistics. In contrast, the high flux of the synchrotron sources allowed detector slit size no larger than 1~$\times$~1~mm$^{2}$. Typical counting times for diffuse scattering measurements was 1 sec/point at APS and  $\leq$10 sec/point at NSLS. High energy harmonics in the incident x-ray beam for all of our experiments were suppressed to the level that did not cause any observable data contamination. Distribution of the scattered intensity in the reciprocal space was measured by fully automated computer-controlled reciprocal space scans. 2D mesh scans allowed for direct  point-by-point reciprocal space mapping. The interval between the scan points was \ensuremath{\sim}0.01-0.02 reciprocal lattice units (r.l.u.), where 1 r.l.u. is ${2 \pi}/a \approx$1.55~\AA$^{-1}$ and $a\approx$4.04~{\AA} is PMN's lattice constant. 2D diffuse scattering measurements were performed 0.1 r.l.u. away  from the Bragg peak centers.

\section{Results and Discussion}

\begin{figure}
     \includegraphics[height=.27\textheight]{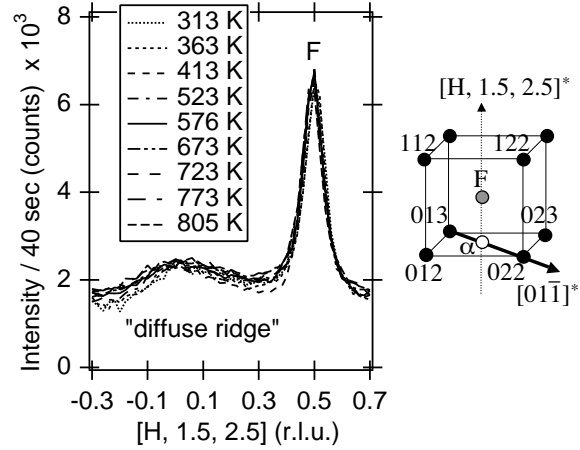}
  \caption{Linear [H,1.5,2.5]* scan in the reciprocal unit cell (shown on the right) measured in PMN above T$_f$ at 313-805 K. Fundamental Bragg peaks are at the corners of the cube. Two superlattice peaks are on the face-centered and the body centered positions (H=0 and H=0.5) along the scan direction.}
  \label{alphahot}
  \end{figure}

Figure~\ref{alphahot} depicts linear $[ H, 1.5, 2.5 ]^*$ reciprocal scan measured in PMN above T$_{f}$\ensuremath{\approx}220~K. Its direction in the reciprocal space can be found on the sketch of the reciprocal unit cell, where the corners are fundamental $Pm\bar{3}m$ Bragg peaks. Two Brillouin zone boundary peaks at H=0.5 and H=0 (F and "diffuse ridge") occupy body-centered and face-centered positions in the reciprocal cell, respectively. These peaks are more than \ensuremath{\sim}10$^{8}$ times weaker and \ensuremath{\sim}100 times broader than Bragg peaks. Figure~\ref{alphahot}  shows that both peaks are temperature independent even above T$_{d}\approx$620 K, where PNRs start to nucleate  \cite{AVT.218}.  Diffuse scattering distribution in the surrounding reciprocal space, required for unambiguous  peak interpretation, was measured using planar 2D mesh scans. 

\begin{figure}
     \includegraphics[height=.26\textheight]{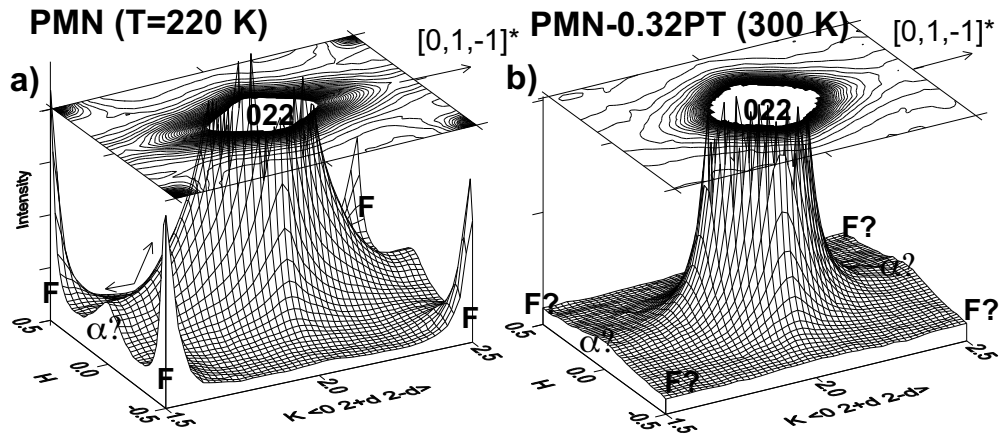}
  \caption{Diffuse scattering intensity maps of the shaded area in Fig.~\ref{recipcube1} near 022 Bragg peak (q>0.1 r.l.u.): (a) PMN  at 220 K ; (b) PMN-0.32PT at 300 K.  Horizontal axes correspond to [110]* and [011]* reciprocal lattice directions. Diffuse scattering intensity is plotted on the vertical linear scale.}
  \label{pmn220k}
  \end{figure}

\begin{figure}
          \includegraphics[height=.17\textheight]{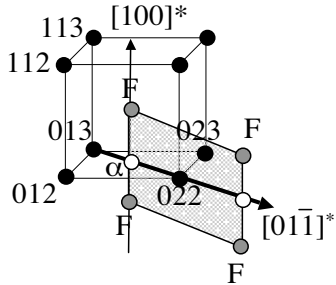}       
  \caption{ Reciprocal unit cell: Bragg peaks (cube corners); F and $\alpha$ spots (body-centered and face-centered positions). Mesh scans in  Fig.~\ref{pmn220k} were measured in the shaded plane.}
  \label{recipcube1}
  \end{figure}

Figure~\ref{pmn220k}  shows reciprocal mesh scans about (022) Bragg peak in PMN at 220~K and in PMN-0.32PT at 300~K in the plane shaded in Fig.~\ref{recipcube1}. Note that the linear H scan in Fig.~\ref{alphahot} is also a part of this plane. (022) Bragg peak is outside of vertical scale and only the diffuse scattering tails $\geq$0.1 (r.l.u.) away from the peak center are visible.  Moreover, FWHM of the (022) Bragg peak is \ensuremath{\sim}0.001 (r.l.u.), which is in agreement with its long-range order nature. Positions of the gray circles in Fig.~\ref{recipcube1} correspond to superlattice peaks at four corners of Fig. ~\ref{pmn220k}(a), which are referred as F spots in the literature \cite{AVT.54}. Origin of the F spots is commonly attributed to chemical nanodomains (CND) produced by $\langle111\rangle$ correlated Nb/Mg short-range order \cite{AVT.54, AVT.184}. Measured FWHM\ensuremath{\approx}0.75 (r.l.u.) of the F spots in Fig.~\ref{pmn220k}(a) and Fig.~\ref{alphahot} corresponds to expected for CND $\sim$50 {\AA} correlation length, which was obtained using  the well-known Scherrer equation \cite{AVT.1002}. 

Existence of the face-centered 1/2(0kl) superlattice peaks \citep{AVT.54, AVT.184, AVT.201, AVT.27, AVT.228}, commonly referred as ${\alpha}$ spots in the literature, is questionable from Fig.~\ref{pmn220k}(a), since diffraction peaks produced by any kind of real space 3D correlations must exhibit a peak cross section in any direction in the reciprocal space \cite{AVT.1002}. However,  there is no clear evidence for any cross section of the ${\alpha}$ spot (marked with a question mark) along $[01\bar{1}]$* direction (marked with a double ended arrow). It appears that a peak cross section at the position of the ${\alpha}$ spot in Fig 2(a) along H ([100]*) direction is produced predominantly by the tail of the diffuse scattering ridge extending from the (022) Bragg peak. The FWHM\ensuremath{\approx}0.25~(r.l.u.) of this cross section corresponds to \ensuremath{\sim}15 {\AA} correlation length at 220~K. This same peak cross section is labeled "diffuse ridge" in Fig.~\ref{alphahot} when measured above 220~K. These results can explain observation of the ${\alpha}$ spots in PMN above \ensuremath{\sim}220~K \cite{ AVT.54,AVT.184,AVT.201,AVT.228} in some cases due to measuring "diffuse ridge" cross sections instead of  actual superlattice peaks from 3D correlations. In fact, the first report of the  ${\alpha}$ spot with synchrotron x-rays was also presented using  linear [100]* reciprocal scan, which cuts through $[01\bar{1}]$* diffuse scattering ridge \cite{AVT.184}.

Fig.~\ref{pmn220k}(b) demonstrates that no superlattice peaks of any kind were found in PMN-0.32PT crystal at 300 K from the mesh scan performed with an identical experimental setup. No superlattice reflections were observed for this composition at any temperature down to 10 K. Anisotropy and strength of the diffuse scattering also appears to be much weaker than in the case of pure PMN. 

It is well know that diffuse scattering in PMN at $q\ll$0.1 (r.l.u.) from the Brillouin zone centers exhibits strong temperature dependence \cite{AVT.184,AVT.1000,AVT.25}. In contrast, temperature independent (see Fig.~\ref{alphahot}), up to 800~K, diffuse scattering at zone boundaries (q=0.5~r.l.u.) suggests that diffuse scattering observed in different parts of the reciprocal space comes from different origins. Moreover, temperature independent $\langle1\bar{1}0\rangle$* diffuse scattering ridges were reported to exist in related Pb(Sc$_{1/2}$Nb$_{1/2}$)O$_{3}$ (PSN) relaxor \cite{AVT.28}. The origin of the ridges has been attributed to static or dynamic displacements in $\{1\bar{1}0\}$ planes without (or with weak) correlation between the planes along  $\langle1\bar{1}0\rangle$ directions. Presence of the ${\alpha}$ spots  in PSN was attributed to existence of the linear anti-ferroelectric chains \cite{AVT.5}. However, displacement correlation lengths were determined from the ${\alpha}$ spots without taking into account cross sections of the overlapping $\langle1\bar{1}0\rangle$ diffuse ridges.

It appears that  ${\alpha}$ spots in PMN can be studied as separate  entities from the diffuse ridges only below 220~K \cite{AVT.198}. Ambiguity related to the detection and interpretation of the ${\alpha}$ spots above T$_{f}$ \cite{AVT.54,AVT.184,AVT.201,AVT.228} can be resolved by proper separation of the temperature dependent ${\alpha}$ spots from temperature independent anisotropic $\langle01\bar{1}\rangle$* diffuse ridges at zone boundaries \cite{AVT.1003}. Therefore,  only $\langle01\bar{1}\rangle$* reciprocal scans were used to obtain widths and integrated areas of the ${\alpha}$ spots needed for structure factor calculations.

\begin{figure}
     \includegraphics[height=0.3\textheight]{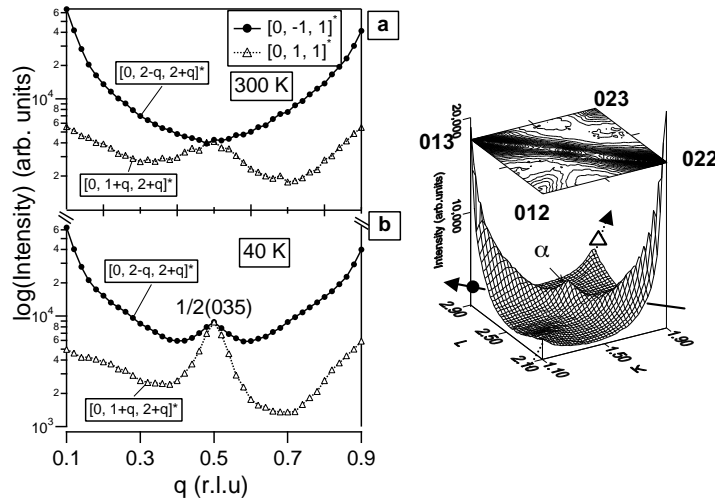}
  \caption{ Proof of coexistence of the $\alpha$ spot and diffuse ridge in PMN below T$_f$ from reciprocal scans along $[0\bar{1}1]$* (solid circles) and  $[011]$* (empty triangles) measured at: (a) 300 K; (b)  40 K. Directions of the scans are indicated with two arrows on the mesh scan (right hand side of the graph) obtained at 45 K.}
  \label{alphacold}
  \end{figure}

Right hand side of Fig.~\ref{alphacold} depicts another 2D mesh scan obtained in PMN at 45~K. It was measured 0.1 (r.l.u.) away from four Bragg peaks on the bottom face of the cube in Fig.~\ref{recipcube1}. Note that ${\alpha}$ spot, which was not pronounced at 220 K along $\langle01\bar{1}\rangle$* in Fig.~\ref{pmn220k}(a), is now clearly visible along this same direction on top of the diffuse scattering ridge connecting (013) and (022) Bragg peaks. Left hand side of Fig.~\ref{alphacold} shows two pairs of linear scans extracted from the similar mesh scans at 300 K and 40 K along two mutually perpendicular $[0\bar{1}1]$* and [011]* directions.  Directions of these scans are indicated on the mesh scan with two diagonal arrows labeled by the same type of markers. Figure~\ref{alphacold}(b) proves existence of the 1/2(035) ${\alpha}$ spot at 40~K. At the same time Fig.~\ref{alphacold}(a) reveals a broad cross section of the $[0\bar{1}1]$* diffuse ridge at 300 K (similar to the one in Fig.~\ref{alphahot}) and also proves the absence of the ${\alpha}$ spot above T$_{f}$. Note that data in Fig.~\ref{alphacold}(a) and (b) are plotted on the same vertical scale, which proves that intensity of the diffuse ridge (solid circles) does not exhibit any pronounced changes below or above T$_{f}$.

\begin{figure}
\resizebox{.96\columnwidth}{!}
    {\includegraphics{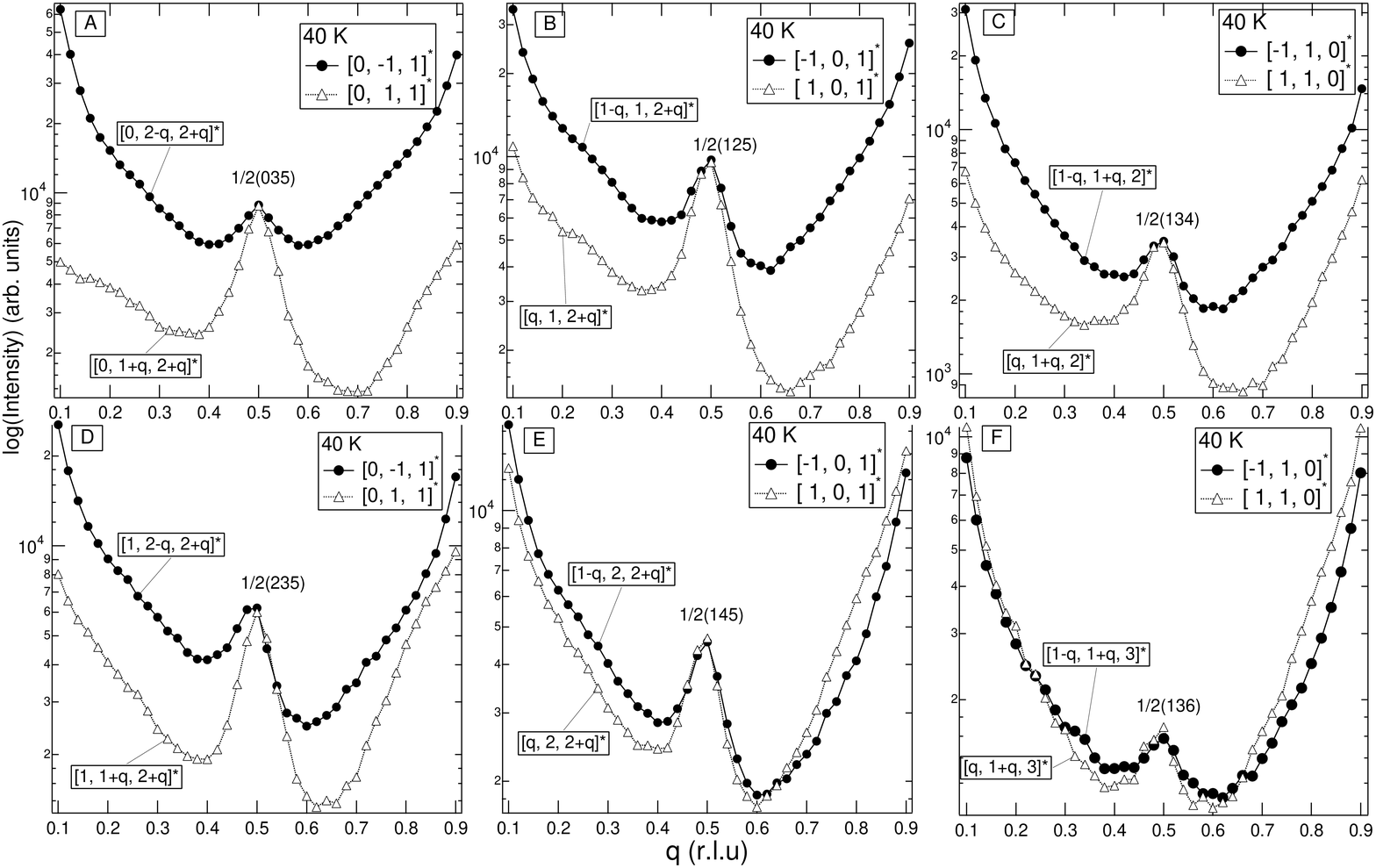}}
  \caption{Direct proof of the $\alpha$ spot presence below T${_f}$ on all six cubic faces in  Fig.~\ref{recipcube1}. Directions of the $\langle01\bar{1}\rangle^*$ diffuse ridges are along the curves plotted with solid circles. }
  \label{sixalpha}
  \end{figure}

	Similar low T measurements, which are presented in Fig.~\ref{sixalpha}, were performed on all six faces of the cube shown in Fig.~\ref{recipcube1}. These data clearly demonstrate that large errors would occur if integrated intensities of the ${\alpha}$ spots are extracted from the empty triangle scans rather than from the solid circle ones. The relative intensities of the $\langle01\bar{1}\rangle$* ridges are different and appear to be correlated with the structure factors of the Bragg peaks that they connect. For example, 1/2(145) and 1/2(136) $\alpha$ spots in Fig.~\ref{sixalpha}(e) and (f), which are located between relatively weak Bragg peaks, are the least affected by the anisotropy of the diffuse scattering background.

\begin{figure}
\resizebox{1.\columnwidth}{!}
     {\includegraphics{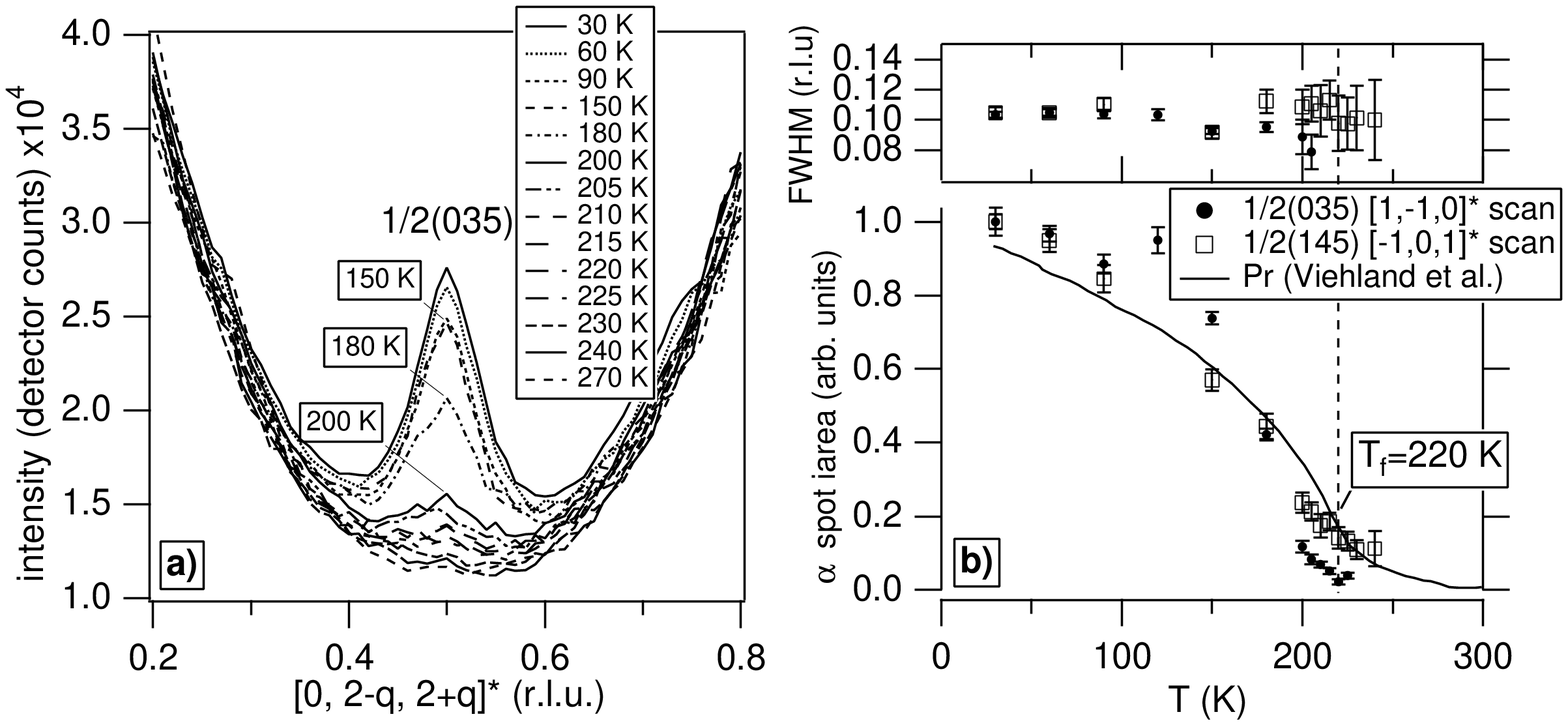}}
  \caption{ (a) Temperature dependence of the 1/2(035) $\alpha$ spot along the ridge in PMN; (b) integrated intensity and FWHM extracted from 1/2(035) and 1/2(145)  $\alpha$ spots.  Normalized remanent polarization P$_r$ is from Ref.\cite{AVT.49} as a guide to the eye.}
  \label{alpha035t}
  \end{figure}

	Figure~\ref{alpha035t}(a) presents detailed temperature dependence of the 1/2(035) $\alpha$ spot measured along $\langle01\bar{1}\rangle$* diffuse ridge without electric field. Integrated intensity and FWHM values together with the ones obtained from the 1/2(145) $\alpha$ spot are plotted vs. temperature in Fig.~\ref{alpha035t}(b). Normalized remanent polarization P$_{r}$ digitized from Ref. \cite{AVT.49} is also shown on the same graph as a guide to the eye. Remarkably, significant reduction in the intensity of the $\alpha$ spots on heating occurs near the phenomenological freezing phase transition T$_{f}$, which is not accompanied by any kind of Bragg peak splitting \cite{AVT.52,AVT.51} in the absence of applied electric field \cite{AVT.126,AVT.21}. This does not contradict average cubic structure, since FWHM$\approx$0.1 (r.l.u.) of the $\alpha$ spots in Fig.~\ref{alpha035t} corresponds to ${\sim}$30 {\AA} correlation length. It also appears to be temperature independent below T$_{f}$ within experimental errors. 

	Similarly, in PMN-0.06PT \ensuremath{\alpha} spots  appear below 265 K \cite{AVT.198}, which corresponds to the freezing temperature for this composition.  However, in PMN-0.1PT only $\langle01\bar{1}\rangle$* diffuse ridges were present at 100 K \cite{AVT.1003}, while  F and \ensuremath{\alpha} spots were not resolvable from the diffuse scattering background for this and higher PT compositions.

\begin{figure}
    {\includegraphics [width=0.9\textwidth] {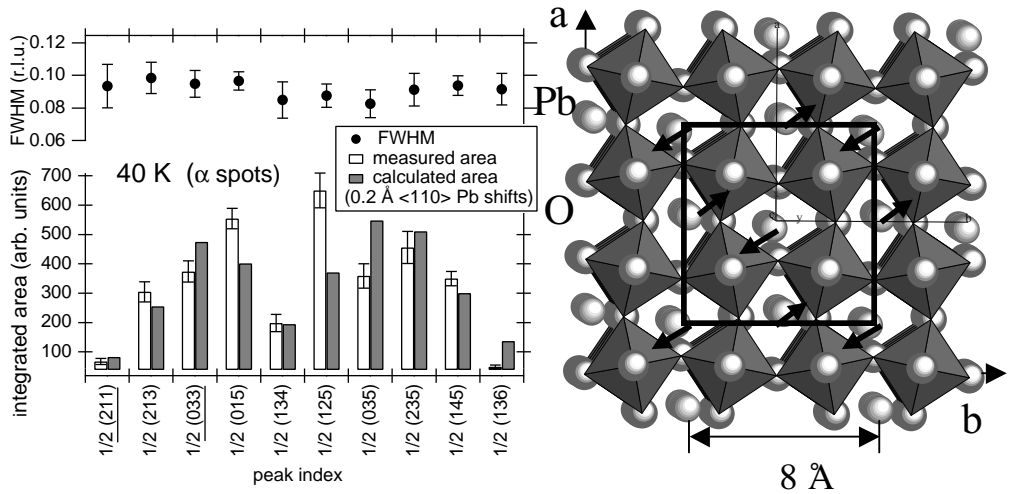}}
  \caption{ Observed and calculated $\alpha$ spot integrated intensities using 0.2 {\AA} Pb$^{2+}$ displacements correlated along equivalent $\langle011\rangle$ directions within 30 \AA~ nanodomains. Shown [110] Pb displacements double unit cell along $a$ and $b$ directions. In-phase oxygen octahedra rotations are  along the $c$ axis.}
  \label{fitarea}
  \end{figure}

Measured at 40 K, integrated intensities of the  $\alpha$ spots are represented as white bars in Fig.~\ref{fitarea} after performing standard absorption and Lorentz factor data corrections \cite{AVT.1002}.  All the peaks have the same FWHM regardless of their location from the center of the reciprocal space. This indicates that the width of the \ensuremath{\alpha} spots is primarily defined by the 30 {\AA} size of the corresponding nanodomains. On the contrary, lattice strain peak broadening generally increases away from the center of the reciprocal space \cite{AVT.1002}.

Temperature dependence of the  $\alpha$ spots excludes their chemical origins and supports presence of correlated on the nanometer scale atomic displacements along all symmetry equivalent $\langle011\rangle$ cubic directions, thereby retaining average cubic symmetry of the whole crystal. According to Glazer classification scheme \cite{AVT.95}, $\alpha$ spots may result from in-phase oxygen octahedra rotations, common to perovskites. However, structure factor calculations show that 1/2(033) and 1/2(211) reflections underlined in Fig.~\ref{fitarea} must be systematically absent \cite{AVT.1003}, which indicates that displacements of other atoms must be involved. For example, Pb$^{2+}$ ions with their lone electron pair are known to be displaced from their equilibrium sites \cite{AVT.67,AVT.208}. These displacements will contribute to the structure factor of the $\alpha$ spots only if some fraction of them is correlated in anti-parallel fashion that doubles the unit cell in $\langle011\rangle$ directions within 30 {\AA} nanodomains. The results of the least squares fit with only two variable parameters are presented as gray bars in Fig.~\ref{fitarea}. All reflections were fitted at once. One of the fitting parameters was a magnitude of the Pb displacements \ensuremath{\delta}, and another was common for all peaks intensity scaling factor. Debye-Waller factors were taken from the literature \cite{AVT.67}. Corresponding Pb displacements with \ensuremath{\delta}=0.2 {\AA}, which gave the best fit, are presented on the right hand side of Fig.~\ref{fitarea}.  Structure factor calculations showed, that this pattern would only contribute to the $\alpha$ spots, which have even third Miller index ($l$), such as 1/2(134) and 1/2(136) in  Fig.~\ref{sixalpha}. \ Similar patterns with Pb displacements having 30 {\AA} correlation length along other equivalent $\langle011\rangle$ directions will contribute to reflections with $h$ or $k$ even indices, accordingly.  

	Inclusion of Nb displacements or in-phase oxygen octahedra rotations with angles up to 10$^{\circ}$ only marginally improved the fit. Neutron measurements, which are more sensitive to oxygen, are clearly needed in order to better understand the role of oxygen octahedra rotations. What is important is that oxygen octahedra alone cannot explain $\alpha$ spot structure factor without Pb displacements. Contribution of Pb displacements to the $\alpha$ spots is consistent with differential anomalous factor scattering (DAFS) measurements near Pb L$_{III}$ absorption edge \cite{AVT.1003}. On the contrary, we were not able to find any evidence for Nb contribution from the DAFS measurement near Nb $K$ edge \cite{AVT.1003}.

	Anti-parallel Pb displacements, which give rise to the \ensuremath{\alpha} spots are expected to compete with ferroelectrically active Pb ions, which are believed to be arranged in a parallel fashion \cite{AVT.67,AVT.1000}. Because displacements are correlated only on the short-range scale, existence of the Pb displacements in other than $\langle110\rangle$ directions is also possible. For example, anti-parallel short-range correlated $\langle111\rangle$ Pb Nb displacements can contribute to the structure factor of the F spots \cite{AVT.31,AVT.208,AVT.1003}. Uncorrelated Pb displacements, on the other hand, will only contribute to the uniform diffuse scattering background. 

In a  separate experiment we observed expected \cite{AVT.126,AVT.21} rhombohedral splitting of the Bragg peaks in PMN(111) \cite{AVT.1003} under electric field $\ge$1.8 kV/cm. However, we did not register any changes in either FWHM or intensity of the \ensuremath{\alpha} spots in both field cooled and zero field cooled regimes below T$_{f}$. This fact indicates that changes in the average macrostructure have no effect on either size or number of the nanoregions that give rise to the \ensuremath{\alpha} spots. On the contrary, the number of PNRs was shown to increase on cooling near T$_{f}$ from the recent electric-field-induced polarization measurements \cite{AVT.224}.

\section{Conclusions}

Possible misinterpretation of the $\alpha$ spots above 220~K for cross sections of the temperature independent  $\langle01\bar{1}\rangle$* diffuse scattering ridges, was addressed and studied in this work. Separation of the diffuse scattering from the $\alpha$ spot superlattice peaks was achieved along $\langle01\bar{1}\rangle$* by reciprocal space mapping utilizing synchrotron x-ray radiation. Correlation length, obtained from the width of the $\alpha$ spots, is only $\sim$30~{\AA}, which defines the average size of producing these peaks nanodomains. These nanodomains are formed by short-range correlated anti-parallel Pb displacements along equivalent $\langle110\rangle$  directions with a magnitude of \ensuremath{\sim}0.2 {\AA} based upon the structure factor calculations. 
Fluctuations created by these locally correlated displacements are different from chemical nanodomains (CND) and ferroelectric polar nanoregions (PNR); they constitute a new type of fluctuations with anti-ferroelectric type displacement ordering (AFR) based on anti-parallel nature of the Pb displacements.

Freezing phase transition has been identified in PMN near T$_{f}\approx$220 K from the temperature dependence of the integrated intensity of the $\alpha$ spots. Significant enhancement in the intensity of the $\alpha$ spots below T$_{f}$ we attribute to increase in a total number of AFRs, which average size ($\sim$30 {\AA}) remains constant down to the lowest measured temperature of 10 K. Nothing can be said about dynamics of these fluctuations, since interaction time between electrons and x-rays during the scattering process is $\sim$10$^{-15}$~sec.

Interestingly, temperature dependence of the \ensuremath{\alpha} spots in related Pb(In$_{1/2}$Nb$_{1/2}$)O$_{3}$ (PIN) variable order relaxor was shown to be correlated with long-range macroscopic anti-ferroelectric phase transition in the case of the fully ordered PIN \cite{AVT.27}.
Competition between randomly occurring anti-ferroelectric and ferroelectric fluctuations may be responsible for the relaxor behavior in PMN.

\begin{theacknowledgments}
We would like to thank Dr. Zschack from UNICAT ID-33 beamline at Advanced Photon Source (APS), Argonne National Laboratory and Dr. Erlich from MATRIX X-18A beamline at National Synchrotron 
Light Source (NSLS), Brookhaven National Laboratory for technical assistance during the experiments.

Authors also would like to thank Dr. Colla and Prof. Feigelson 
for providing good quality PMN single crystals and enlightening 
discussions.

This research is based upon work supported by the U.S. Department 
of Energy, Division of Materials Sciences under award No. DEFG02-96ER45439 
and by the state of Illinois IBHE HECA NWU A207 grant, though 
the Frederick Seitz Materials Research Laboratory at the University 
of Illinois at Urbana-Champaign.

National Synchrotron Light Source (NSLS) is supported by the 
U.S. Department of Energy under Contract No. DE-AC02-76CH00016. 
Use of the Advanced Photon Source was supported by the U.S. Department 
of Energy, Basic Energy Sciences, Office of Science, under Contract 
No. W-31-109-Eng-38.

\end{theacknowledgments}


\bibliographystyle{aipproc}   

\bibliography{ferro03a}

\IfFileExists{\jobname.bbl}{}
 {\typeout{}
  \typeout{******************************************}
  \typeout{** Please run "bibtex \jobname" to optain}
  \typeout{** the bibliography and then re-run LaTeX}
  \typeout{** twice to fix the references!}
  \typeout{******************************************}
  \typeout{}
 }

\end{document}